\documentclass[fleqn,10pt]{wlscirep}
\usepackage{array}
\usepackage{tabularx}
\title{Interplay of charge density wave and multiband superconductivity in 2$H$-Pd$_x$TaSe$_2$}

\author[*]{D. Bhoi, S. Khim, W. Nam, B. S. Lee, Chanhee Kim, B.-G. Jeon, B. H. Min, S. Park, and Kee Hoon Kim}
\affil[1]{Center for Novel States of Complex Materials Research and Institute of Applied Physics, Department of
Physics and Astronomy, Seoul National University, Seoul 151-747, Republic of Korea}

\affil[*]{Corresponding author khkim@phya.snu.ac.kr}
\begin{abstract}
2$H$-TaSe$_2$ has been one of unique transition metal dichalcogenides exhibiting several phase transitions due to a delicate balance among competing electronic ground states. An unusual metallic state at high-$T$ is sequentially followed by an incommensurate charge density wave (ICDW) state at $\approx$ 122 K and a commensurate charge density wave (CCDW) state at $\approx$ 90 K, and superconductivity at $T_{\rm{C}}\sim$0.14 K. Upon systematic intercalation of Pd ions into TaSe$_2$, we find that CCDW order is destabilized more rapidly than ICDW to indicate a hidden quantum phase transition point at $x$$\sim$0.09-0.10. Moreover, $T_{\rm{C}}$ shows a dramatic enhancement up to 3.3 K at $x$ = 0.08, $\sim$24 times of $T_{\rm{C}}$ in 2$H$-TaSe$_2$, in proportional to the density of states $N(E_F)$. Investigations of upper critical fields $H_{c2}$ in single crystals reveal evidences of multiband superconductivity as temperature-dependent anisotropy factor $\gamma_H$ = $H_{c2}^{ab}$/$H_{c2}^{c}$, quasi-linear increase of $H_{c2}^{c}(T)$, and an upward, positive-curvature in $H_{c2}^{ab}(T)$ near $T_{\rm{C}}$. Furthermore, analysis of temperature-dependent electronic specific heat corroborates the presence of multiple superconducting gaps. Based on above findings and electronic phase diagram vs $x$, we propose that the increase of $N(E_F)$ and effective electron-phonon coupling in the vicinity of CDW quantum phase transition should be a key to the large enhancement of $T_{\rm{C}}$ in Pd$_x$TaSe$_2$.
\end{abstract}
\begin{document}
\newcolumntype{L}[1]{>{\raggedright\arraybackslash}p{#1}}
\newcolumntype{C}[1]{>{\centering\arraybackslash}p{#1}}
\newcolumntype{R}[1]{>{\raggedleft\arraybackslash}p{#1}}
\flushbottom
\maketitle
\thispagestyle{empty}
\section*{Introduction}

Transition metal dichalcogenides (TMDCs) have been extensively studied for decades due to their rich electronic properties resulting from their low dimensionality. These systems have the layered structure and share the formula MX$_2$, where M is a transition metal atom (M = Ti, Zr, Hf, V, Nb, Ta, Mo, W and Re), and X is a chalcogen atom (X = S, Se, and Te) \cite{will1,will2,ross}. Each layer consists of a hexagonal transition metal sheet sandwiched by two similar chalcogen sheets and the layers are coupled to each other by weak van der Waals force. Within the layers, they form strongly bonded, two dimensional X-X layers while M has either trigonal prismatic or octahedral coordination with X. Many of these compounds exhibit charge-density wave (CDW) and some of them manifest the competition/coexistence between CDW and superconductivity \cite{will1,will2,ross}. It is common in many of these compounds that superconductivity emerges in the vicinity of the CDW quantum phase transition induced by intercalation of variety of elements into the van der Waal gaps \cite{gamble1,gamble2,fang1,mora1,mora2,wagner} and by the application of pressure \cite{kums,sipos} or gate voltage to the pristine compound \cite{yijun}. The overall electronic phase diagram as a function of doping, gate voltage or pressure are thus analogous to those of high-$T_{\rm{C}}$ cuprates, iron based superconductors and heavy fermion materials, suggesting possible role of the CDW quantum critical point on the creation of superconductivity. However, exact physical origin for stabilizing the superconductivity over the CDW state has not been well understood and should be further clarified in each TMDC system.\\

The 2$H$-polymorph of tantalum diselenide (TaSe$_2$) is yet another renowned CDW system, in which an unusual metallic state at high temperatures enters into an incommensurate-charge-density wave (ICDW) phase at $T_{\rm{ICDW}} \approx $122 K, followed by a first order ``lock-in" transition to a commensurate charge density-wave (CCDW) phase at $T_{\rm{CCDW}} \approx$ 90 K \cite{mon1,mon2,flem}. Angle-resolved photoemission spectroscopy (ARPES) and optical spectroscopy measurements have suggested the presence of pseudogaps for $T > T_{\rm{ICDW}}$ similar to that observed in high-$T_{\rm{C}}$ cuprates \cite{boris,vescoli,ruzicka}. Interestingly, in the temperature region, resistivity is known to change quite linearly with temperature \cite{vescoli,ruzicka}. Upon entering the ICDW phase, these pseudogaps become larger and finally evolves into real band-gaps in the CCDW phase. At the same time, two hole-like circular pockets around $\Gamma$ and $K$ points and one electron-like `dogbone' centered on $M$ point of the Brillouin zone (BZ)\cite{boris,ross} show drastic reconstruction at $T=T_{\rm{CCDW}}$, resulting in three times smaller BZ, characterized by doubly degenerate circular pockets and small rounded triangles in the $K$ and $M$ points of the new BZ, respectively \cite{boris}. While the origin of the CDW instability in 2$H$-TMDCs has been a topic of debate for many years, most recent studies indicate that the CDW formation in 2$H$-TMDCs is driven by periodic lattice distortion induced by strong momentum-dependent electron-phonon coupling rather than by the Peierls-type mechanism \cite{weber,dai2,johan,johan1,liu}. \\

Inside the CCDW state, the 2$H$-TaSe$_2$ exhibits yet another phase transition into a superconductor at $T_{\rm{C}}\sim$ 0.14 K. This superconductivity arises by the conventional electron-phonon coupling. It has been known that chemical substitution of Te for Se and doping of Mo or W for Ta changed the 2$H$-TaSe$_2$ into 3$R$ polytypes \cite{luo1,luo2}. In this case, an optimal $T_{\rm{C}}$ is found to be 2.4 K, of which origin has been attributed to the three-layer stacking sequence of 3$R$-TaSe$_2$ compared to two-layer stacking sequence in 2$H$-TaSe$_2$. Meanwhile, the superconductivity can be further modulated by a specific doping of Fe or Ni into 2$H$-TaSe$_2$ within the same 2$H$-structure \cite{li,whit,dai}.
In this case, only intercalation of Ni was found to show enhancement of $T_{\rm{C}}$ up to 2.7 K \cite{li}, suggesting that not only structural motif but also carrier modulation might be important for superconductivity. Therefore, to clarify the exact roles of structure and carriers on the delicate balance between the CDW and superconductivity, systematic doping/intercalation studies in each structural unit become necessary.

In this report, we show that intercalation of Pd ion into 2$H$-TaSe$_2$ in a broad doping ranges without altering the 2$H$ structure enhances $T_{\rm{C}}$ up to 3.3 K, the highest observable $T_{\rm{C}}$ among the TaSe$_2$ polytypes. With intercalation of Pd, the CCDW phase of 2$H$-TaSe$_2$ is suppressed more rapidly than the ICDW transition while the $T_{\rm{C}}$ increases continuously to form a dome-shape phase diagram and an optimal $T_{\rm{C}}\sim$ 3.3 K for $x$ = 0.08-0.09. We find that $T_{\rm{C}}(x)$ increases in proportional to the density of states estimated from the Sommerfeld coefficient. Measurements of upper critical fields and specific heat below $T_{\rm{C}}(x)$ show clear evidences of multiple superconducting energy gaps. These observations and comparison with other TMDCs uncover that the increase of density of states and thus electron-phonon coupling constant, regardless how it is driven by, e.g., structural modification or by carrier doping, could be important for the creation of multiband BCS superconductivity in the vicinity of the quantum phase transition of the CCDW state.

\section*{Results}
\subsection*{Powder x-ray diffraction}
\begin{figure}[h]
  \centering
  \includegraphics[width=0.75\textwidth]{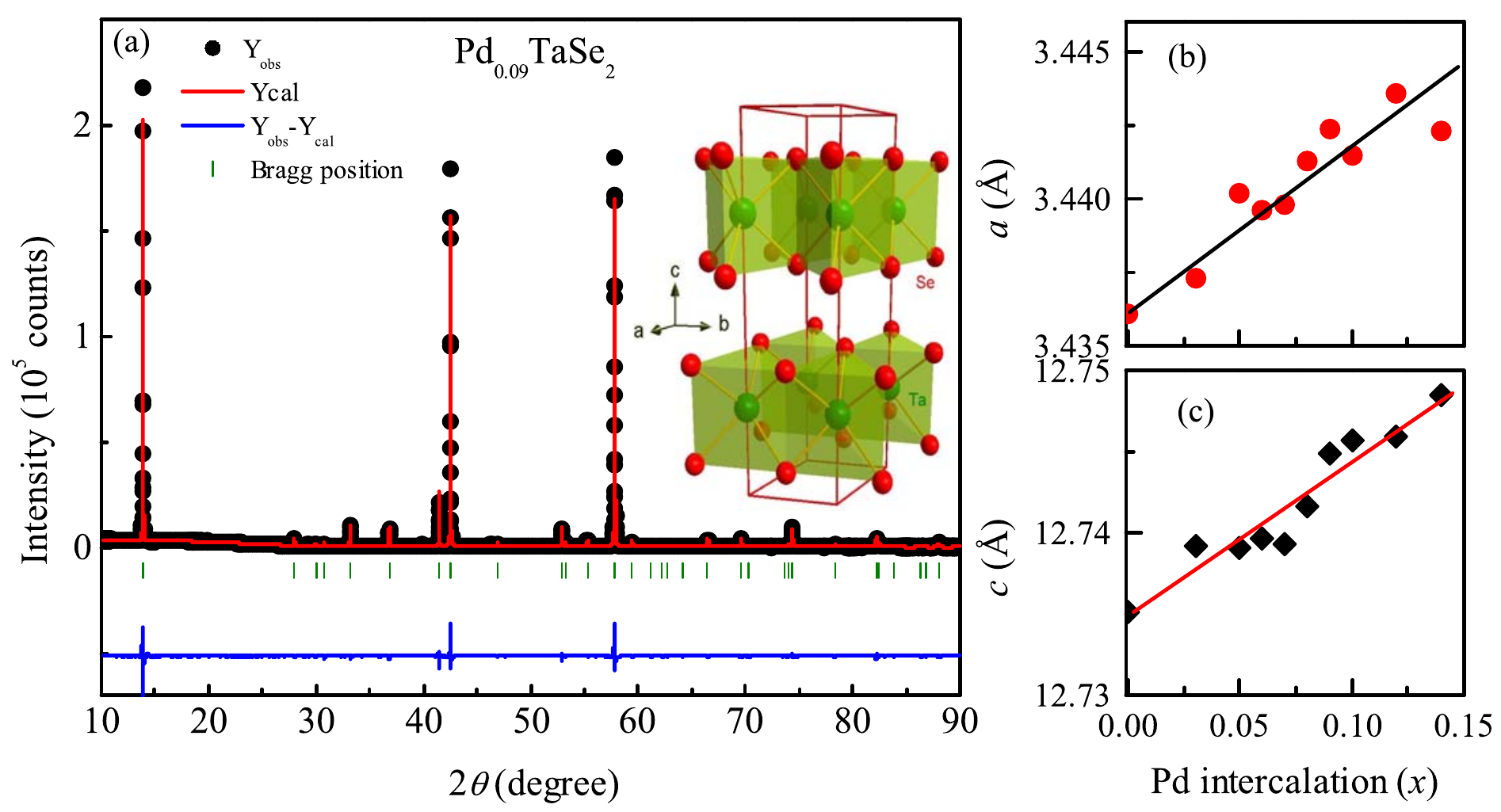}\\
  \caption{(a) Powder x-ray diffraction pattern of the Pd$_{0.09}$TaSe$_2$ sample and quantitative refinement confirming the 2$H$-polytype structure. Inset shows the crystal structure of the 2$H$-TaSe$_2$. Variation of the (b) $a$-axis and (c) $c$-axis lattice parameters with Pd content ($x$). Standard deviations of the cell parameters are smaller than the points.}\label{Fig1}
\end{figure}
Powder x-ray diffraction patterns of the Pd$_x$TaSe$_2$ samples in Fig.\ref{Fig1} indicate that all the peaks can be indexed with the structure of 2$H$-TaSe$_2$. Hence, the process of Pd intercalation does not change the structure of the mother compound. The refinement of the diffraction pattern was used to extract lattice parameters ($a$ and $c$) at room temperature, as summarized in Fig.\ref{Fig1}(b, c). Intercalation of atoms into the van der Waal gaps usually leads to expansion of a unit cell. With increasing Pd content, both $a$- and $c$-values increase systematically with $x$. The observed trend is similar to that observed in 1$T$-Cu$_x$TiSe$_2$ \cite{mora1}, 1$T$-Pd$_x$TiSe$_2$ \cite{mora2}, and 2$H$-Cu$_x$TaS$_2$ \cite{wagner}. The $a$- ($c$-) value of PdSe$_2$ is larger (smaller) than that of 2$H$-TaSe$_2$ \cite{gron}. Hence, if Pd ions are partly substituting Ta ions, TaSe$_2$ would have exhibited the increase of $a$ and decrease of $c$, which is in contrast to our results. Thus, the substitution of Pd for Ta doesn't seem to happen, presumably due to a smaller ion size of Pd ($r_{\rm{Pd}^{+1}}$ = 59 pm) than Ta ($r_{\rm{Ta}^{+4}}$ = 68 pm). All this information supports that Pd ions are mostly intercalated into the layers with the van-der-Waals bonding.\\
\subsection*{Resistivity and magnetization}
\begin{figure}[h]
  \centering
  \includegraphics[width=0.75\textwidth]{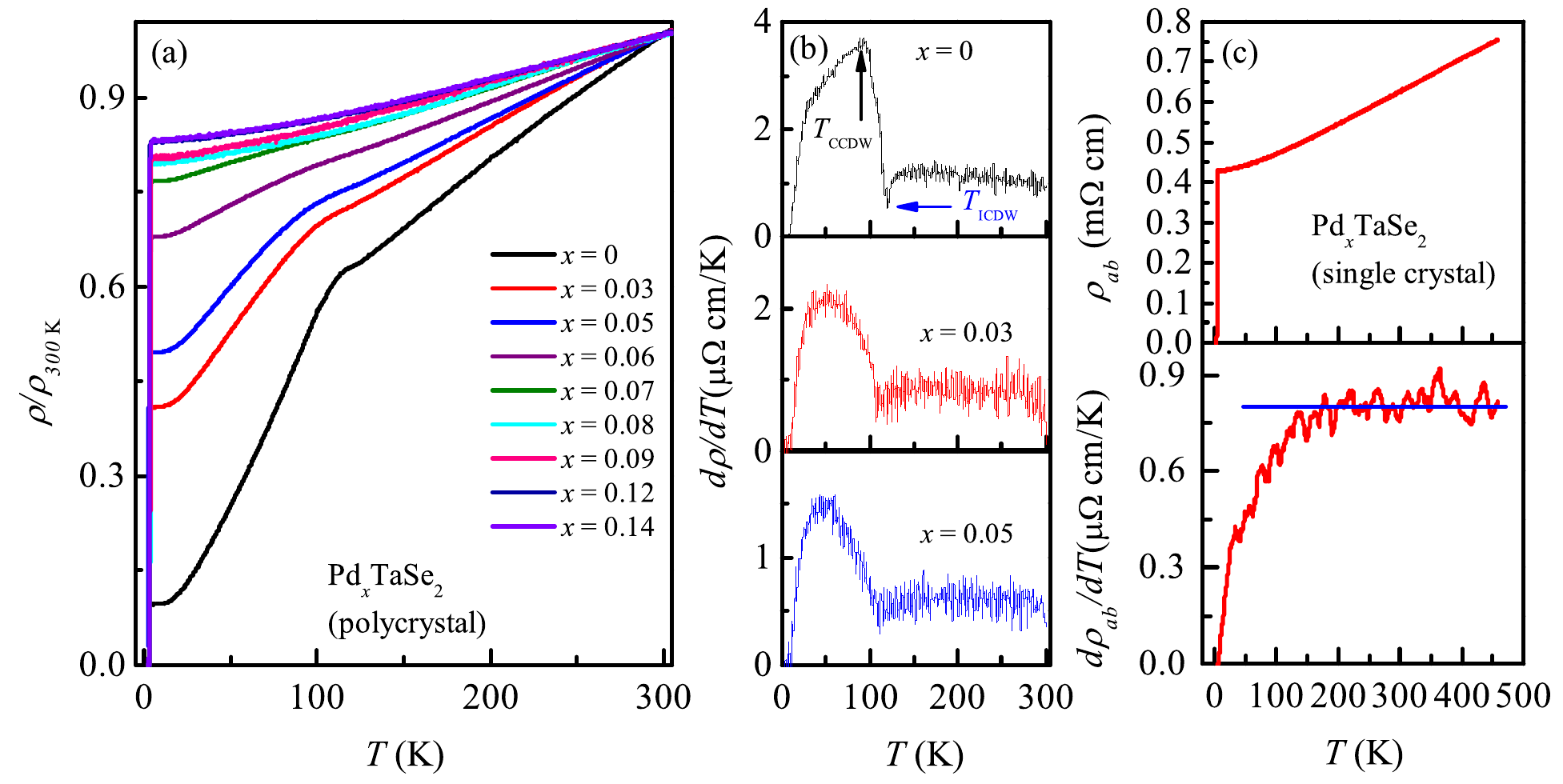}\\
  \caption{(a) Temperature-dependent resistivity normalized to the value at 300 K, $\rho/\rho_{300 K}$, of the Pd$_x$TaSe$_2$ polycrystalline samples for 0$<x<$0.14. (b) $d\rho/dT$ curves of selected samples. Arrow marks at the minimum and maximum of $d\rho/dT$ indicate the ICDW and CCDW transition temperatures, respectively. (c) Top: temperature-dependent in-plane resistivity $\rho_{ab}$ of a Pd$_x$TaSe$_2$ single crystal up to 460 K, with $T_{\rm{c}} \sim$ 3.2 K. Bottom: The $d\rho_{ab}/dT$ curve exhibiting almost constant behavior demonstrates the $T$-linear dependence of resistivity at high temperatures at least up to 460 K.}\label{Fig2}
\end{figure}
In general, near the CDW transition temperature, resistivity and magnetic susceptibility curves exhibit an anomaly due to the opening of a partial or full gap at the Fermi energy ($E_{\rm{F}}$), which is often accompanied by the simultaneous decrease of the Fermi surface area and the variation of density of states $N$($E_{\rm{F}}$). Hence, the evolution of CDW states with $x$, can be traced from the temperature-dependent resistivity and susceptibility measurements. Figure \ref{Fig2}(a) presents the resistivity normalized by the value at 300 K ($\rho/\rho_{300 K}$). The undoped 2$H$-TaSe$_2$, $\rho$($T$) shows the metallic behavior near room temperature to low temperatures till it exhibits a hump near 115 K due to the formation of ICDW-state, being consistent with the former studies \cite{will1,vescoli}. Below this transition, resistivity steeply decreases further to exhibit more metallic behavior. The two CDW transitions can be indeed identified from the $d\rho/dT$ curve, developing a minima near $T_{\rm{ICCDW}}\approx$ 120 K and a maxima near $T_{\rm{ICCDW}}\approx$ 90 K. The transition temperatures thus estimated are in good agreement with those obtained from neutron diffraction studies \cite{mon1,mon2,flem}.

With increase of Pd content, the residual resistivity ratio (RRR) of Pd$_x$TaSe$_2$ compounds decreases systematically compared to the undoped compound, indicating that the compounds thus evolve into a bad metal. Meanwhile, $\rho_{300 K}$ of all the samples was located around 0.2-0.6 m$\Omega$ cm without showing systematic trend, due to the polycrystalline nature of the samples. With increase of $x$, the anomalies in $d\rho/dT$ (Fig.$\ref{Fig2}$(b)) becomes broadened. Yet, the minima of $d\rho/dT$ curve could be identified up to $x$=0.10 and the maxima of $d\rho/dT$ curve up to $x$=0.07, allowing the determination of $T_{\rm{ICCDW}}$ and $T_{\rm{CCDW}}$. It is also clear from the $d\rho/dT$ curve in Fig.$\ref{Fig2}$(b) that the maximum points indicating $T_{\rm{CCDW}}$ decreases more rapidly with $x$ while the minima points for $T_{\rm{ICDW}}$ decreases slightly. These behavior is summarized later in the electronic phase diagram.
\begin{figure}[h]
   \centering
  \includegraphics[width=0.55\textwidth]{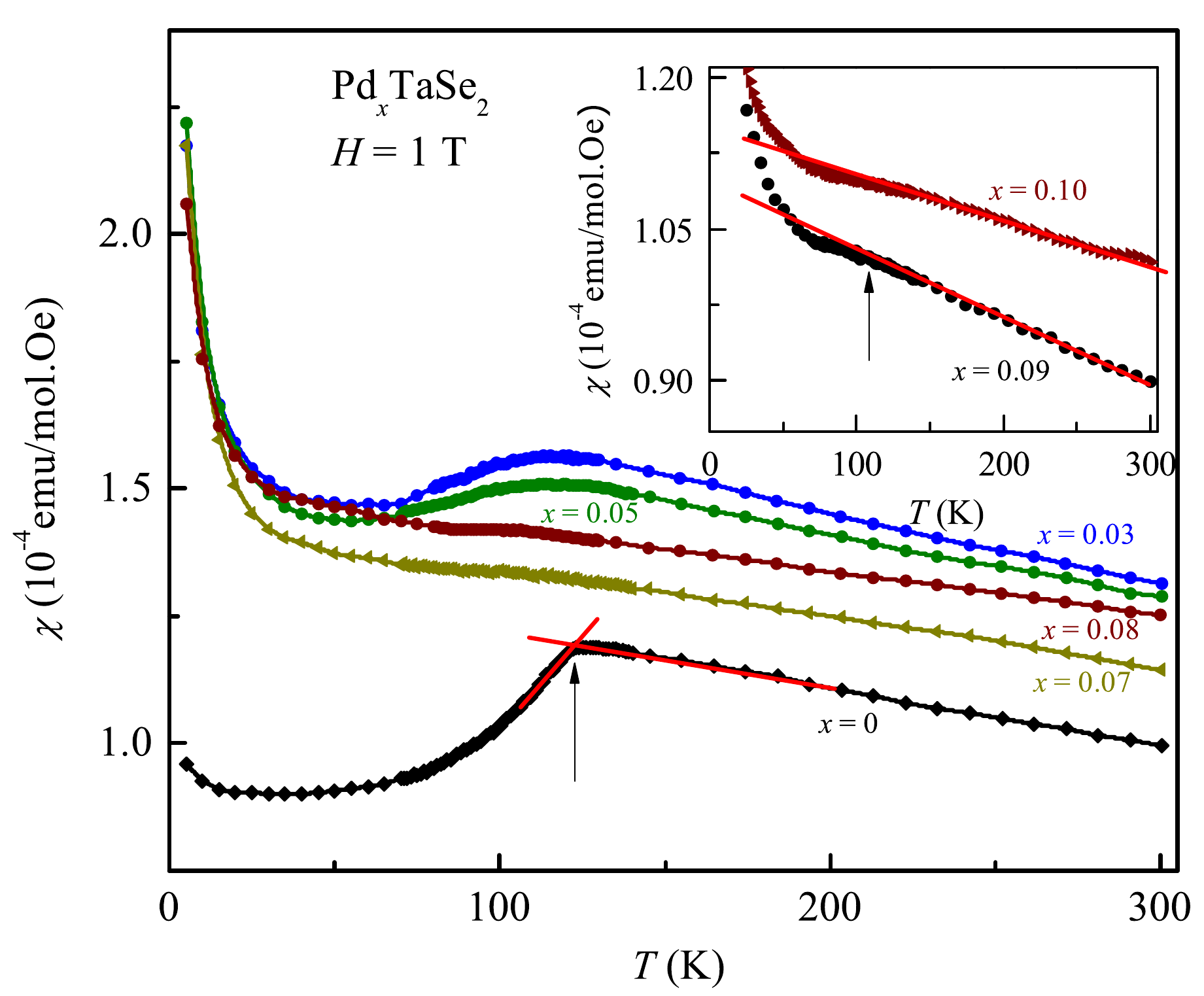}\\
  \caption{Temperature-dependence of normal state susceptibility $(\chi)$ of Pd$_x$TaSe$_2$ samples measured at constant applied field $H$ = 1 T for 0$\leq x \leq$ 0.08. Inset shows the $\chi$($T$)curves for $x$ = 0.09 and 0.10. The red solid lines illustrate how the ICDW transition temperatures have been determined. The arrows mark the $T_{\rm{ICDW}}$ for the undoped and $x$ = 0.09 (inset) samples.}\label{Fig3}
\end{figure}
For all $x$, $\rho$($T$) below 30 K follows a power law behavior ($T^\alpha$) with $\alpha\approx 2$, being consistent with the Fermi liquid behavior. On the other hand, roughly above 140 K (i,e at $T > T_{\rm{ICDW}}$), where a pseudogap behavior was found in the mother compound \cite{vescoli,ruzicka}, it is interesting to note that $\rho$ increases quite linearly with temperature as evidenced from the almost constant behavior in the $d\rho/dT$ curve (Fig.\ref{Fig2}(b)). To further corroborates the intrinsic nature of this phenomenon, we have measured the in-plane resistivity, $\rho_{ab}$, of a Pd$_{x}$TaSe$_2$ single crystals ($T_{\rm{C}} \sim$ 3.2 K) up to 460 K as shown in Fig.\ref{Fig2}(c). The $\rho_{ab}$ curve indeed shows the linear $T$-dependence at least up to 460 K as evidenced by almost constant behavior of the $d\rho_{ab}/dT$ curve. Note that the measurement was performed under Ar atmosphere to avoid the oxidation of sample while the transport data became unstable above 460 K, possibly due to unknown changes in the sample. The broad $T$-range of observing linear $T$-dependence (from $\approx$140 K to $\approx$460 K) is reminiscent of that observed in high $T_{\rm{C}}$ cuprates. Moreover, the phenomenon itself that 2$H$-TaSe$_2$ develops the pseudo-gap above $T_{\rm{ICCDW}}$ seems also similar to those observed in the high $T_{\rm C}$ cuprates \cite{vescoli,ruzicka}. In conventional metals, the linear increase of $\rho(T)$ can be realized in a temperature window above the Debye temperature due to the electron-phonon scattering, according to the Bloch-Gr\"{u}neisen theory \cite{bg}. However, once the mean free path $l$ at high temperatures reaches the interatomic spacing $a$, the resistivity should start to show saturation. This crossover resistivity is known as the Ioffe-Regel (IR) limit \cite{mott,ioffe,gunar}. In a quasi 2D materials with cylindrical Fermi surface, $\rho$ should saturate at $\rho_{IR}^{2D}$ = $2\pi\hbar d/e^{2}k_{F}l$, where $d$ is interlayer spacing, $k_F$ is Fermi wave vector \cite{gunar}. Taking $d$ = 6.73 $\AA$, half of the $c$-value, and $l\simeq a$ for the Pd$_x$TaSe$_2$ crystal, $\rho_{IR}^{2D}$ = 699 $\mu\Omega$ cm. It is obvious from Fig.\ref{Fig2}(c) that $\rho_{ab}$ continues to increase linearly at least up to 460 K clearly above 390 K where it should have saturated according to the IR limit. Thus, the linear increase of $\rho_{ab}$ curve cannot be simply understood by the conventional electron-phonon scattering scenario. Even in high $T_{\rm C}$ cuprates, the IR criterion is also violated and $\rho_{ab}$ continues to increases linearly as high as 1100 K as a result of strong electronic correlation effect \cite{gunar}. Therefore, the origin of the linear resistivity in the Pd$_{x}$TaSe$_2$ single crystal could be worth being further explored, related with the similar conundrum existing in high $T_{\rm C}$ cuprates.\\

Figure \ref{Fig3} shows dc magnetic susceptibility $(\chi)$ of Pd$_x$TaSe$_2$ samples in the normal-state for different $x$. A drop in the $(\chi)$ curve of pure $2H$-TaSe$_2$ is seen at $T_{\rm{ICCDW}}\approx$ 120 K, in agreement with the resistivity data. $T_{\rm{ICDW}}$ could be also determined from the onset of the susceptibility drops as shown in the fig. \ref{Fig3}. A Curie-Weiss-like tail also appears in the low temperature, possibly due to a small local moment induced by Pd intercalation. Appearance of a similar Curie-Weiss-like tail has also been observed in the Cu and Pd intercalated system \cite{mora1,mora2}. The sharpness of the drop decreases with increasing amount of Pd, and $T_{\rm{ICDW}}$ transition decreases to 107 K for $x$ = 0.10. Due to the Curie-Weiss increase at low temperatures, the ICDW transition could not be identified in the $\chi(T)$ for  $x>$0.10.\\

\begin{figure}[h]
   \centering
  \includegraphics[width=0.75\textwidth]{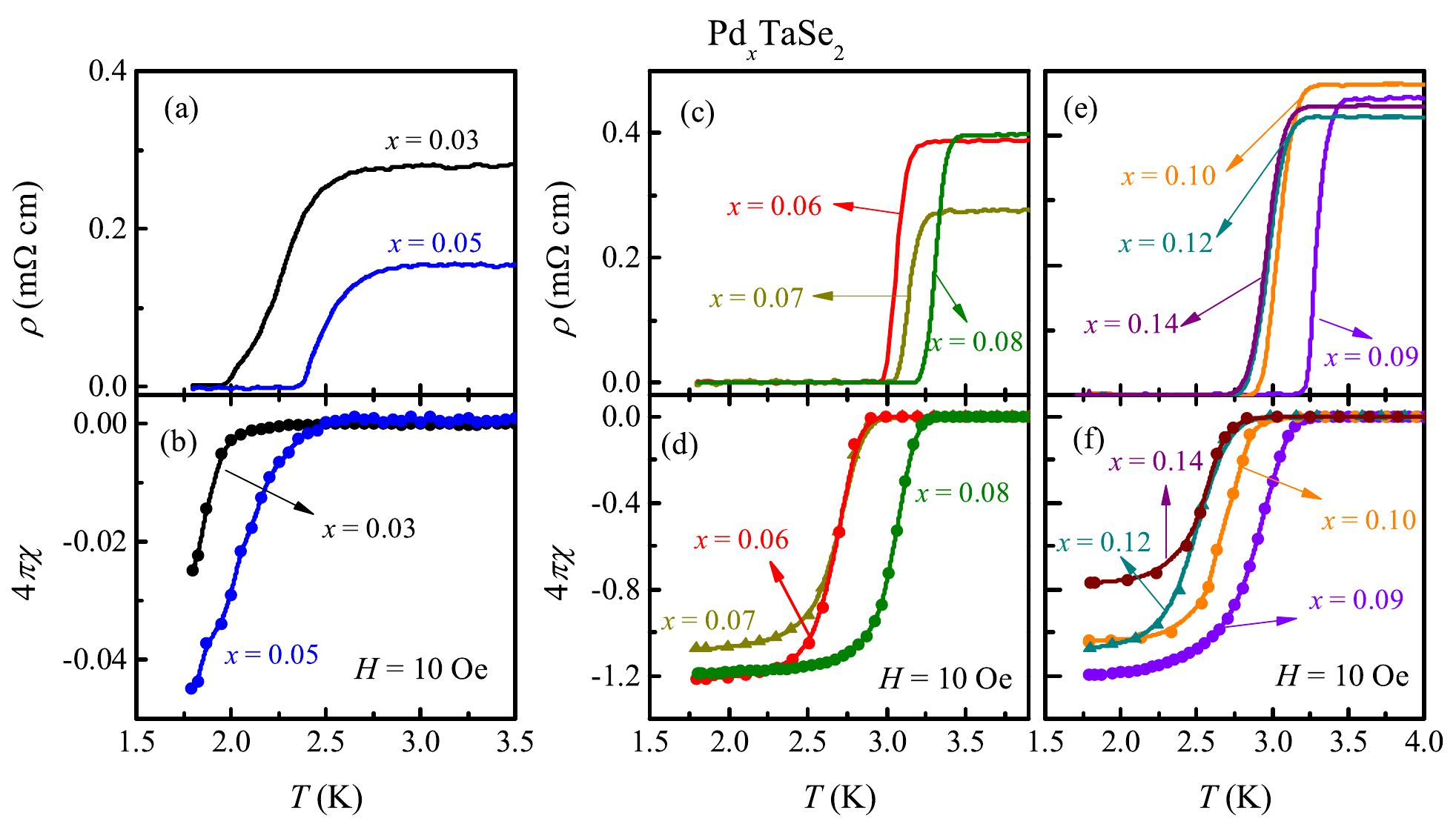}\\
  \caption{Low temperature behavior of zero-field resistivity $\rho$ (a) for $x$ = 0.03 and 0.05, (c) for 0.06 $\leq x \leq $0.08, and (e) for $x$ = 0.09, 0.10, 0.12 and 0.14. Temperature dependence of zero-field cooled susceptibility indicating the superconducting volume fraction (-4$\pi\chi$) measured at $H$ = 10 Oe (b) for $x$ = 0.03 and 0.05, (d) for 0.06 $\leq x \leq $0.08, and (f) for $x$ = 0.09, 0.10, 0.12 and 0.14.}\label{Fig4}
\end{figure}

We found that as the CCDW state is suppressed with increasing Pd content, the superconducting transition temperature increases continuously. This behavior is illustrated in the low-temperature $\rho(T)$ and zero field cooled $\chi(T)$ data as shown in Fig.\ref{Fig4}. For all $x$, the superconducting transition are clearly observed in resistivity while the onset of diamagnetic behavior is also observed in $\chi(T)$. With a small amount of Pd ($x$ = 0.03), $T_{\rm C}$ is enhanced more than 17 times than that of the pure compound. With further increase of Pd content, $T_{\rm C}$ increases systematically and attains a maximum of about 3.3 K for $x$ = 0.09 and then decreases to 2.83 K for $x$ = 0.14. For $x \leq$0.05, the width of superconducting transition in $\rho$ becomes broadened compared to the higher doping samples, possibly because the superconducting volume fraction (-$4\pi\chi$) is less than 10\% at 1.8 K (Fig.\ref{Fig4}(a,b)). For $x\ge$ 0.06, the value of -$4\pi\chi$ at 1.8 K increases significantly and reaches to 120\%, indicating the presence of bulk superconductivity. The shielding fraction larger than 100\% is likely to come from the demagnetization factor of the specimen. For $x>$ 0.09, -$4\pi\chi$ decreases again with increase of $x$ to reach 77\% for $x$ =0.14.\\
\subsection*{Upper critical fields}
\begin{figure}[h]
 \centering
  \includegraphics[width=0.85\textwidth]{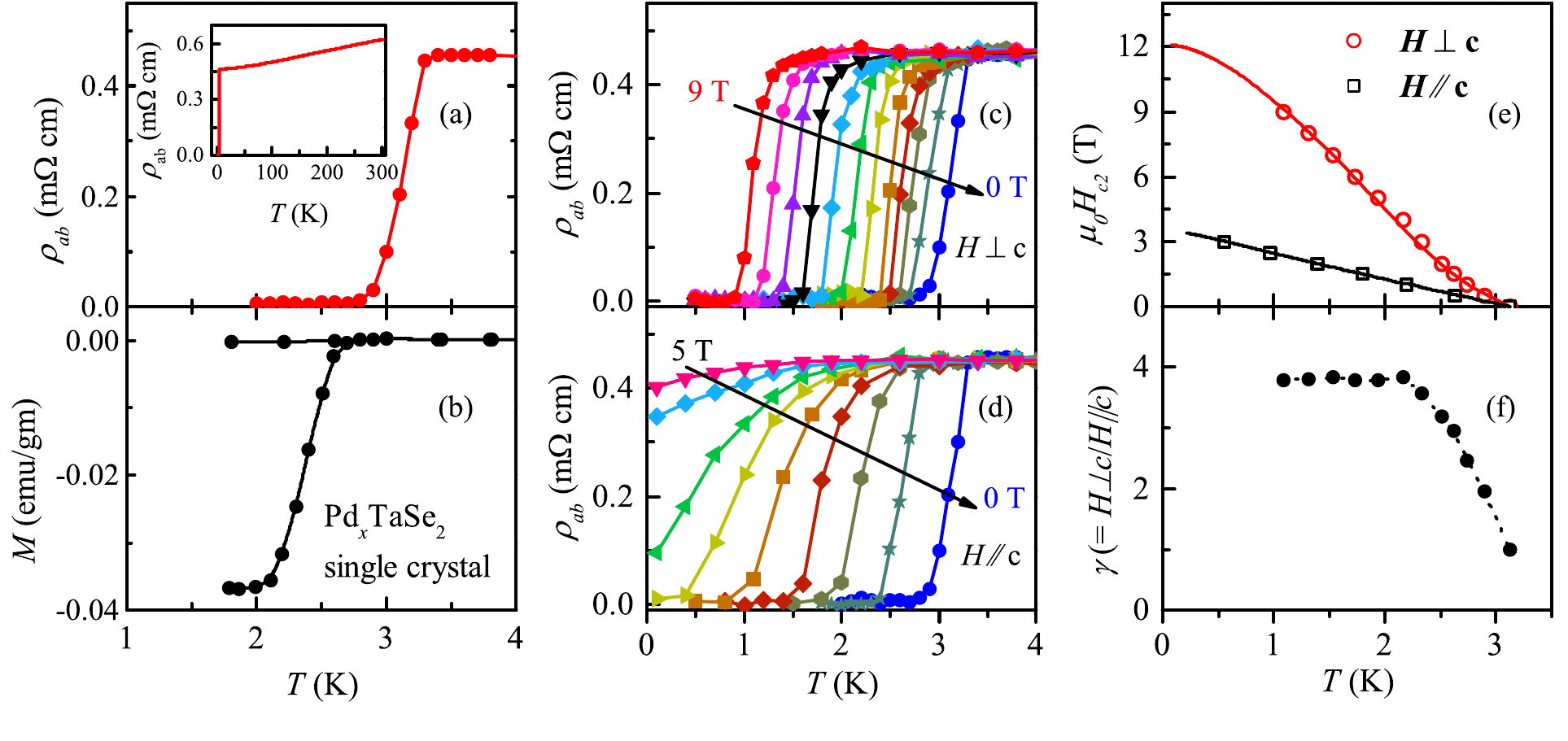}\\
\caption{(a) Zero-field resistivity and (b) dc magnetization at $H$ = 10 Oe for the Pd$_x$TaSe$_2$ single crystal. Inset of (a) shows the normal state resistivity of the crystals. Magnetic field dependence of the resistive transition of Pd$_x$TaSe$_2$ single crystal for (c) $H \parallel ab$ and (d) $H \parallel c$. (e) Temperature dependence of the upper critical fields determined by the 50\% of the normal state resistivity criterion. The solid lines show the fit to the two-band model based on equation \ref{multi}. (f) Temperature dependence of the anisotropy factor $\gamma_H$ = $H_{c2}^{ab}$/$H_{c2}^{c}$ and the dotted line is a guide to the eye.}\label{Fig5}
\end{figure}
For better understanding of the superconducting states, we have measured the temperature-dependence and anisotropy of the upper critical fields, $H_{c2}$, of a Pd$_{x}$TaSe$_2$ single crystal in a dilution refrigerator. Figure \ref{Fig5}(a) and (b) show the $T$-dependence of the zero-field $\rho$ and dc magnetization at $H$ = 10 Oe of the same Pd$_{x}$TaSe$_2$ single crystal used for $H_{c2}$ measurements. In Fig. \ref{Fig5}(c) and (d), we have plotted the magnetic field dependence of $\rho_{ab}$ with $H$ applied parallel and perpendicular to the $ab$-plane. It is clear that for $H \parallel ab$, the resistive curve shifts parallel down towards the low temperature with increase of $H$. For $H \parallel c$, the resistive transition broadens significantly and superconductivity is completely suppressed below 4 T. Figure \ref{Fig5}(e) summarizes the temperature-dependence of estimated $H_{c2}$ for both field directions, i.e. $H_{c2}^{ab}$ and $H_{c2}^{c}$, determined by the criterion of the 50\% of normal state resistivity. It is noteworthy that $H_{c2}^{c}$ increases linearly down to the lowest measured temperature without any sign of saturation. This is in sharp contrast to what is seen in isotropic, single-band BCS superconductors in which $H_{c2}(T)$ follows a linear dependence for temperatures close to $T_{\rm{C}}$, followed by a saturation, or a concave downward curvature at low temperature. Such linear temperature-dependence has been observed in the Fe-based superconductors \cite{khim1} and quasi-one dimensional superconductors \cite{khim,zhang}, indicating the multi-band effects.\\

The evidence of the multiband effect becomes more clear from the temperature dependence of $H_{c2}^{ab}$ and the anisotropy factor $\gamma_H$ = $H_{c2}^{ab}$/$H_{c2}^{c}$. The $H_{c2}^{ab}(T)$ curve exhibits a positive curvature close to $T_{\rm{C}}$ and then increases linearly with decreasing temperature. A positive curvature in the $H_{c2}(T)$ curve has been often observed in numerous multi-gap superconductors such as YNi$_2$B$_2$C, MgB$_2$, NbSe$_2$ and NbS$_2$ \cite{shu,sudr,sudr1,tiss} when their various FSs have different electron-phonon coupling parameters and Fermi velocities. The $\gamma_H(T)$, shown in Fig \ref{Fig5}(f), also supports the multiband effects. For $T$ close to $T_{\rm{C}}$, $\gamma_H\sim$ 1.96, and it increases with decreasing temperature down to 2 K before it becomes saturated. This is quite similar to that observed in MgB$_2$ \cite{gure}, which could be understood as an interplay between two major gaps with temperature; the gap in a cylindrical $\sigma$ band dominates low-$T$ behavior of $H_{c2}$, leading to a large anisotropy, while at $T \simeq T_c$, the gap in the $\pi$ bands play more important role, strongly reducing the anisotropy.\\

In order to have more quantitative understanding, we have analyzed the data using the two-band model for $H_{c2}$, which takes into account the orbital and Zeeman pair breaking effect in the dirty limit \cite{gure}. Assuming negligible interband scattering, the equation of $H_{c2}(T)$ can be written as
\begin{equation}
a_0[lnt + U(h/t)][lnt + U(\eta h/t )] + a_2[lnt +U(\eta h/t )] + a_1[lnt + U(h/t)]= 0,
\label{multi}
\end{equation}
where $a_0$ = 2($\lambda_{11}\lambda_{22}-\lambda_{12}\lambda_{21}$), $a_1$ = 1+($\lambda_{11}-\lambda_{22}$)/$\lambda_0$, $a_2$ = 1-($\lambda_{11}-\lambda_{22}$)/$\lambda_0$, $\lambda_0$ = [$(\lambda_{11}-\lambda_{22})^2 + 4\lambda_{12}\lambda_{21}]^{1/2}$, $h = H_{c2}D_1/2\phi_0 T$ , $t = T/T_c$, $\eta = D_2/D_1$, and $U(x) = \Psi(x+1/2)-\Psi(x)$. $\Psi(x)$ is the digamma function, $\lambda_{11}$ and $\lambda_{22}$ are the intraband BCS coupling constants, while $\lambda_{12}$ and $\lambda_{21}$ are the interband BCS coupling constants, and $D_1$ and $D_2$ are the in-plane diffusivity of each band.

As shown in Fig. \ref{Fig5}(e), the experimental data can be well described within the framework of the two-band model. For $H_{c2}^{ab}(T)$, the best fit parameters are $\lambda_{11} = \lambda_{22} = 0.8$, $\lambda_{12} = \lambda_{21}= 0.11$, $D_1$ = 0.29 and $\eta = 5$, and for the $H_{c2}^c(T)$, $\lambda_{11} = \lambda_{22} = 0.5$, $\lambda_{12} = \lambda_{21}= 0.45$, $D_1$ = 2.97 and $\eta = 0.142$. The obtained fitting parameters suggests that the intraband coupling is much stronger thant the interband coupling. \\
\subsection*{Heat capacity}
\begin{figure}[h]
  \centering
  \includegraphics[width=0.7\textwidth]{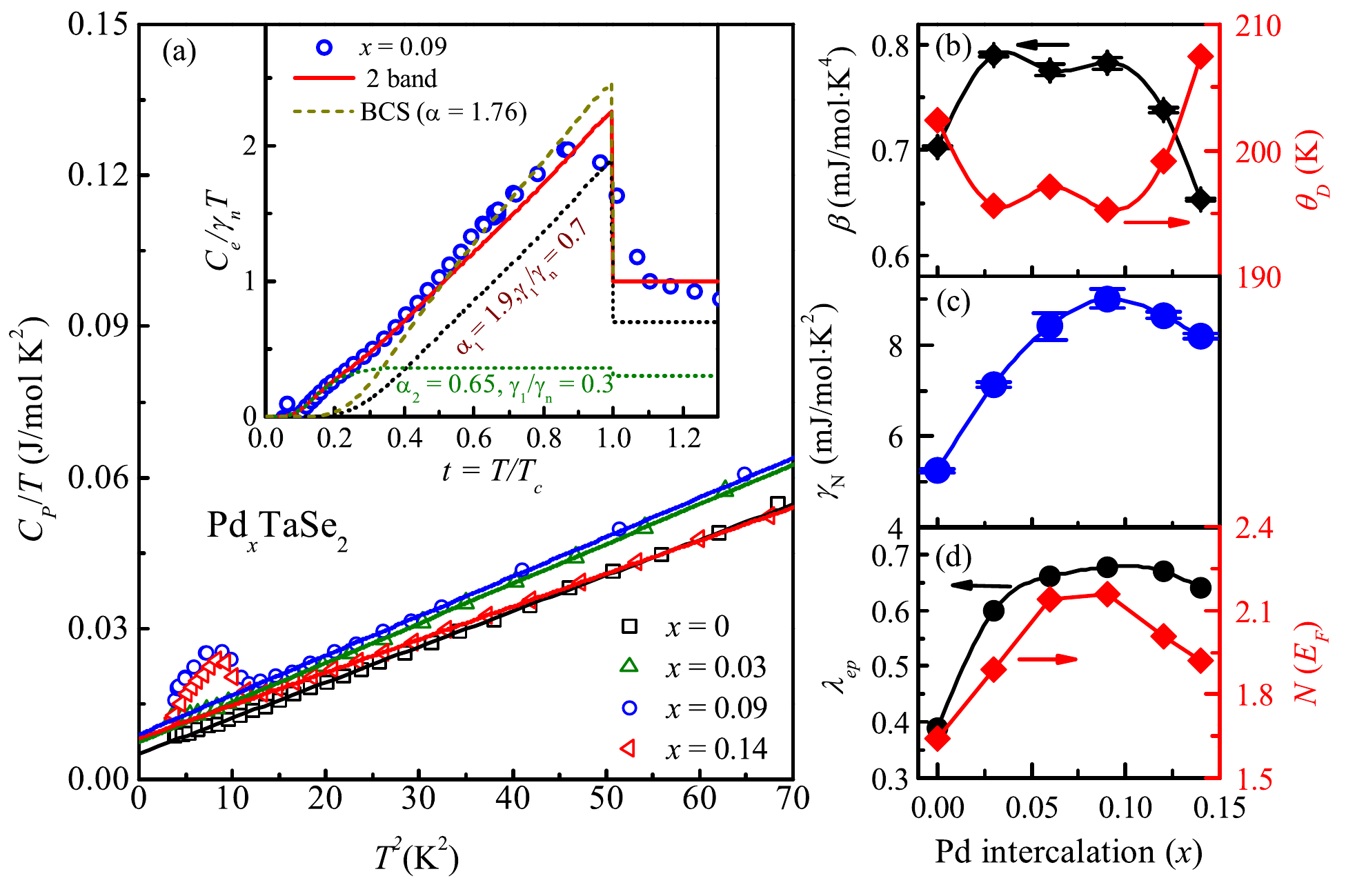}\\
 \caption{(a) The $C_p/T$ vs. $T^2$ curve of the polycrystalline Pd$_x$TaSe$_2$ samples for $x$ = 0, 0.03, 0.09 and 0.14. The solid lines represent the linear fit to the data at temperature between $T_{\rm{C}}$ and 10 K. Insets show the electronic specific heat $C_{el}$ after subtracting phonon and low temperature Schottky contributions for $x$ = 0.09 samples. The dashed (dark yellow) and solid line (red) denote the BCS $\alpha$-model for the one-band ($\alpha$ = 1.76) and the two-band ($\alpha_1$ = 1.9($\gamma_1/\gamma_n$ = 0.7) and $\alpha_1$ = 0.65($\gamma_1/\gamma_n$ = 0.3)), respectively. Variation of (b) $\beta$ and Debye temperature $\Theta_D$ (c) normal state $\gamma_N$ with $x$. (d) Estimated electron-phonon coupling constant $\lambda_{ep}$ using equation \ref{ep} and density of states at the Fermi energy $N(E_{\rm{F}})$ with $x$.}\label{Fig6}
\end{figure}
To obtain further information on the electronic and superconducting properties, we have measured the specific heat $C_p$. Figure \ref{Fig6}(a) displays $C_p/T$ vs $T^2$ curves for selected samples of $x$ = 0.0, 0.03, 0.09, and 0.14. The data for $x$ = 0.09 and 0.14 exhibit clear peaks near $T_{\rm{C}}$, supporting the bulk nature of superconductivity. The normal state specific heat at temperatures between $T_{\rm{C}}$ and 10 K are well described by the relation $C_p =\gamma_{n} T +\beta T^3$, where $\gamma_{n} T$ and $\beta T^3$ term describes the Sommerfeld coefficient and phononic contribution, respectively. The electronic contribution for $x$ = 0.09 was obtained by subtracting the phonon and Schottky contributions from the measured $C_p$ data. In order to extract the Schottky contribution below 0.4 K, we employed a two level energy scheme. Inset of Fig. \ref{Fig6}(a) shows the resultant, normalized electronic specific heat $C_{el}/\gamma_n T$ as function of reduced temperature $t =T/T_c$ for $x$ = 0.09. It is found that $C_{el}/\gamma_n T$ displays a hump-like feature around $t$ = 0.2, which supports the presence of multiple superconducting gaps.

We thus analyzed the $C_p$ data using the BCS $\alpha$-model, which was originally proposed to explain thermodynamic properties of a strongly coupled, single-gap superconductor under a semi-empirical approach \cite{padam}. However, later this model had been generalized to explain the multigap superconductors such as MgB$_2$ \cite{Bouq}. In case of the two gaps, specific heat is determined from the two contributions, i.e. $\alpha_1$ [=$\Delta_1(0)/k_BT_c$] and $\alpha_2$ [=$\Delta_2(0)/k_BT_c$] with their relative weight $\gamma_1/\gamma_n$ and $\gamma_2/\gamma_n$ where $\gamma_1+\gamma_2 = \gamma_n$ and $\Delta(0)$ is the gap value at zero temperature. As evident from the inset of Fig. \ref{Fig6}(a), the calculation for one-band $\alpha$-model with $\alpha$ = 1.76 deviates significantly from the experimental data below $t\approx$0.5. We then calculated $C_{el}/\gamma_n T$, introducing two gaps and their appropriate weights. Values $\alpha_1$ = 1.9 ($\gamma_1/\gamma_n$ = 0.7) and $\alpha_2$ = 0.65 ($\gamma_2/\gamma_n$ = 0.3), produced the closest matching with our experimental data. The inset of Fig.\ref{Fig6}(a) also shows the $C_{el}/\gamma_n T$ vs $t$ plot for an individual $\alpha$ with its weights. The estimated gap values also support the analysis of upper critical fields, thereby confirming the existence of at least two gaps with different magnitudes.\\

By fitting the data at the normal state between $T_{\rm{C}}$ and 10 K, we have evaluated the Sommerfeld coefficient $\gamma_n$ and the phonon specific heat coefficient $\beta$, of which variations with $x$ are summarized in the Fig.\ref{Fig6}(b) and (c). It is surprising to find that $\gamma_n$ systematically increases by more than 80 \% until it reaches a maximum at $x$ = 0.09, and then decreases with further increase of $x$. Note that the evolution of $\gamma_n$ with $x$ is qualitatively similar to that of $T_{\rm{C}}$ with $x$. On the other hand, $\beta$ increases initially, stays almost constant up to $x$ = 0.09, and decreases again about 15 \%. From the $\beta$ value, we can estimate the Debye temperatures by the relation $\Theta_D = (12\pi^{4}nR/5\beta)^{1/3}$, where $n$ is the number of atoms per formula unit ($n$  = 3), and $R$ is the gas constant. For phonon-mediated, one-band superconductors, the electron-phonon coupling constant ($\lambda_{ep}$) can be also estimated from the inverted McMillan equation using $\Theta_D$ and the superconducting transition temperatures \cite{mcmill},
\begin{equation}
\lambda_{ep} = \frac{1.04+\mu^\ast ln\{\frac{\Theta_D}{1.45T_c}\}}{(1-0.62\mu^\ast)ln\{\frac{\Theta_D}{1.45T_c}\}-1.04}.
\label{ep}
\end{equation}
Assuming the Coulomb pseudopotential $\mu^\ast$ to be 0.15 empirically, we have determined the variation of $\lambda_{ep}$ with $x$ as shown in the Fig.\ref{Fig6}(d). Note that the estimated value of $\lambda_{ep} = 0.39$ for the undoped compound is consistent with the reported value in literature \cite{ross}. In case of a multiband superconductor, application of the above equation that has been developed for one band superconductor should have its own limit. However, the obtained $\lambda_{ep}$ could qualitatively reflect the average of the eigenvalues in the electron-phonon interaction matrix \cite{mazin1}.

We found that $\lambda_{ep}$ shows qualitatively similar variation with $T_{\rm{C}}$ but changes within 10 \% except the initial steep increase from 0.39 for $x$ = 0 to 0.60 for $x$ = 0.03. With the value of $\gamma_n$ and $\lambda_{ep}$, the density of states at the Fermi level $N(E_{\rm{F}})$ can be calculated from the equation $N(E_{\rm{F}})$ = $\frac{3}{\pi^2 k_B^2(1+\lambda_{ep})}\gamma_n$. Figure \ref{Fig6}(d) shows the evaluated $N(E_{\rm{F}})$ with $x$. Similar to $\gamma_n$, $N(E_{\rm{F}})$ increases systematically with $x$ from 1.64 states/eV f.u. for $x$ = 0 to 2.16 states/eV f.u. for $x$ = 0.09, and decreases again at higher doping above 0.09. Therefore, it is concluded that the Pd intercalation into the 2$H$-TaSe$_2$ indeed resulted in the maximum values of $\gamma_n$ and $N(E_{\rm{F}})$ near the optimal doping level of $x$=0.09, similar to the evolution of $T_{\rm{C}}$. The increase of $N(E_{\rm{F}})$ as doping is increased toward an optimal doping has been previously found in TaSe$_{2-x}$Te$_x$ too. However, the structure of TaSe$_{2-x}$Te$_x$ has changed from the 2$H$ to 3$R$ structure with Te doping\cite{luo1} so that it was difficult to judge whether the structural change was essential or not to increase of $N(E_{\rm{F}})$ (or $\lambda_{ep}$) and thus $T_{\rm{C}}$ of BCS superconductivity in a broad class of TMDCs. The observation of qualitative proportionality between $N(E_{\rm{F}})$ and $T_{\rm{C}}$ in Pd$_x$TaSe$_2$ with the same structure indicates that the increase of $N(E_{\rm{F}})$ (or $\lambda_{ep}$) becomes more fundamental to increase $T_{\rm{C}}$, regardless of the structural evolution in many TMDCs.\\

\section*{Phase diagram and Discussion}
\begin{figure}[h]
  \centering
  \includegraphics[width=0.6\textwidth]{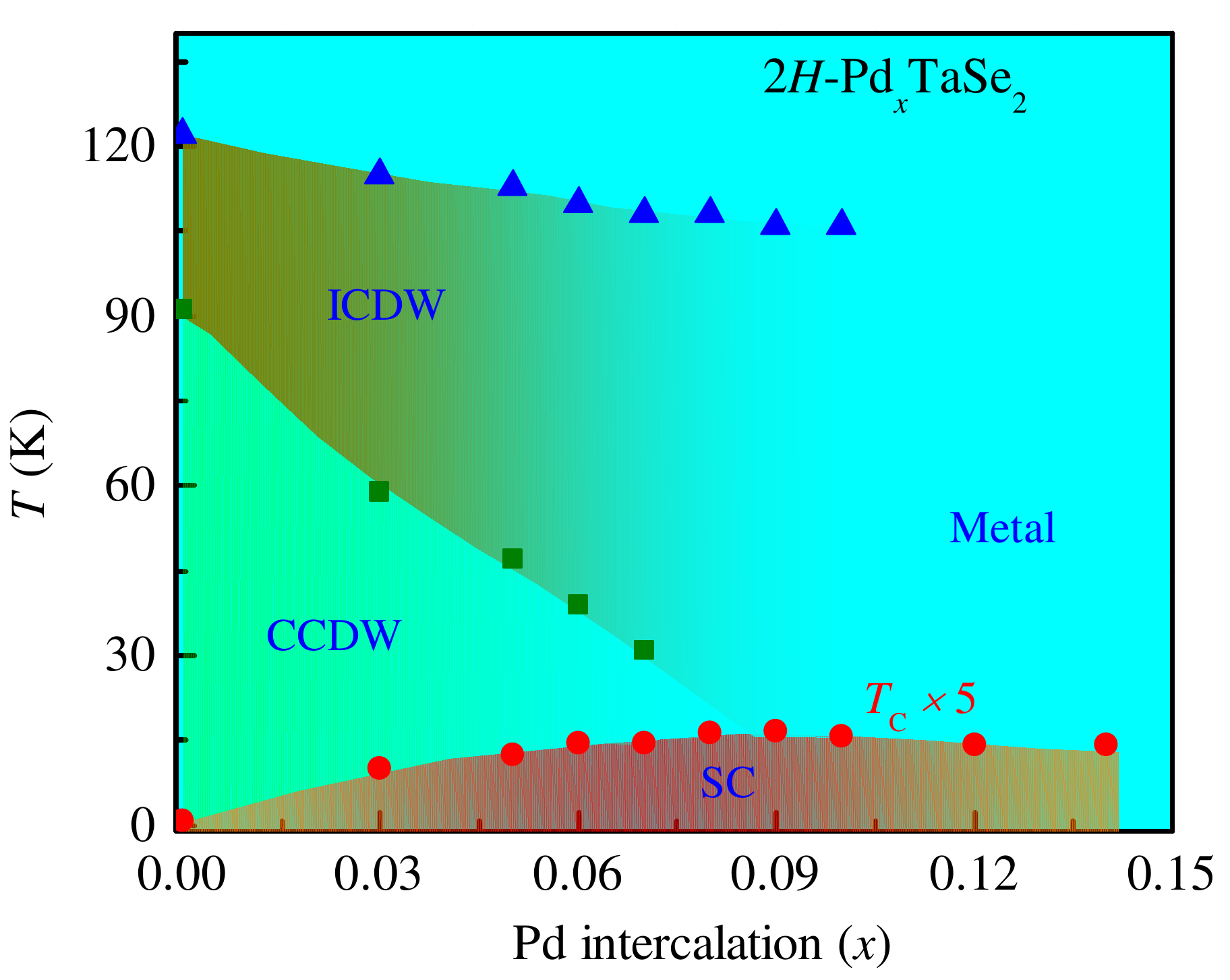}\\
  \caption{Electronic phase diagram of 2$H$-Pd$_x$TaSe$_2$ with Pd intercalation fraction $x$. The blue triangles represent $T_{\rm{ICDW}}$ determined from the normal state susceptibility data, the green squares represent $T_{\rm{CCDW}}$ determined from the maximum of the $d\rho/dT$ curve in Fig.\ref{Fig2} and the red circles represent $T_{\rm{C}}$ determined from the onset of diamagnetic susceptibility in Fig.\ref{Fig4}. For clarity the $T_{\rm{C}}$ value have been multiplied by 5 times.}\label{Fig7}
\end{figure}
We have summarized the electronic phase diagram in the 2$H$-Pd$_x$TaSe$_2$ system as shown in Fig.\ref{Fig7}. With Pd intercalation, the CCDW phase is suppressed dramatically compared to the ICDW phase. As the $T_{CCDW}$ is progressively suppressed, $T_c(x)$ increases systematically. The phase diagram reveals a classic dome shape with an optimal $T_{\rm{C}}$ for $x$ = 0.08-0.09. Most importantly, the collapse of CCDW phase is almost coincident with the doping levels of showing an optimal $T_{\rm{C}}$, supporting clearly the competition between CCDW phase and superconductivity. On the other hand, $T_{ICDW}$ slowly decreases from 122 to 106 K regardless of the superconducting dome, suggesting that there is no direct association with each other.

Related to this, it is worthwhile to discuss the distinct features of the phase diagram of 2$H$-Pd$_x$TaSe$_2$ system with other systems with the same 2$H$ structure. The presence of two CDW transitions is indeed unique in the 2$H$-TaSe$_2$ because all other 2$H$ compounds have shown only one incommensurate CDW transition. For example, both 2$H$-NbSe$_2$ and 2$H$-TaS$_2$ showed only the ICDW phase \cite{will1,will2,mon1}. Furthermore, although either pressure or intercalation of Cu ions suppressed the ICDW phase and enhanced superconductivity in 2$H$-NbSe$_2$ \cite{leroux} and 2$H$-TaS$_2$ \cite{wagner}, respectively, the collapse of the ICDW did not occur at the optimal $T_{\rm{C}}$ in both cases. For 2$H$-NbSe$_2$, the collapse point is located well below optimal $T_{\rm{C}}$ while for 2$H$-TaS$_2$, it is well above. Therefore, direct competition between ICDW and superconductivity doesn't seem to happen in those other compounds with the 2$H$ structure. Consistent with this, it has been recently claimed by theoretical calculations and inelastic x-ray scattering that the electron-phonon coupling of the soft acoustic mode could be mainly responsible for ICDW formation in 2$H$-NbSe$_2$, while the optical phonon mode is coupled to the creation of superconductivity \cite{leroux}. Our observation of very weak dependence of ICDW phase with superconductivity clearly supports that superconductivity is not related to the ICDW phase in 2$H$-Pd$_x$TaSe$_2$ too, being similar to the case in 2$H$-NbSe$_2$ \cite{leroux}. This also implies in 2$H$-TaSe$_2$ that a specific type of electron-phonon coupling required to stabilize the ICDW phase might be also different from that for inducing superconductivity. Furthermore, the strong competition of the CCDW with the superconductivity in turn suggests that a same type of phonon mode might be responsible for stabilizing both the superconductivity and CCDW in 2$H$-TaSe$_2$.

One interesting question to check is whether the increase of $T_{\rm{C}}$ via the increase of $N(E_{\rm{F}})$ and $\lambda_{ep}$ as observed here could be generally applied to other TMDCs even with different structures and dopants. For this, we have compared previous experimental results of $T_{\rm{C}}$, $\gamma_n$, and $\theta_D$, and calculated $N(E_{\rm{F}})$ and $\lambda_{ep}$ in various known TMDCs in the literature as summarized in Table \ref{scp}. To our surprise, we find that although significant variation exists in $N(E_{\rm{F}})$ but not in $\lambda_{ep}$ within the doped 2$H$-TaSe$_2$ and 2$H$-TaS$_2$, $N(E_{\rm{F}})$ is almost proportional to $T_{\rm{C}}$ in each system, regardless of their structural forms. Therefore, $N(E_{\rm{F}})$ seems to be the key factor rather the structural motif for determining $T_{\rm{C}}$ magnitude. At the same time, when $N(E_{\rm{F}})$ is almost same, the variation of $\lambda_{ep}$, albeit small, is still helpful in increasing $T_{c}$. Moreover, between the materials with a different parent compound such as 1$T$-Cu$_x$TiS$_2$ and 2$H$-TaS$_2$, the magnitude of $\lambda_{ep}$ was also important to determine the $T_{\rm{C}}$ magnitude. Based on all these results, we conclude that $T_{\rm{C}}$ is strongly dependent on $N(E_{\rm{F}})$ while it is also affected by $\lambda_{ep}$.

At present, what is a generic origin for the variation of $N(E_{\rm{F}})$ in the Pd$_x$TaSe$_2$ system remains as an open question. One scenario is that $N(E_{\rm{F}})$ increases due to the proximity of the system to the quantum critical point of the CCDW state. If this scenario is true, it is expected that the collapse of multiple Fermi Surface in the CCDW across the quantum critical point could cause the increase of $N(E_{\rm{F}})$. Another scenario is that the Pd intercalation has provided the increase of partial Pd density of state by direct hybridization with the TaSe$_2$ states near the Fermi level or by indirect lattice modulation. It might be worthwhile to check the feasibility of both scenarios in future studies.

In conclusion, we have investigated the effect of Pd intercalation into the 2$H$-TaSe$_2$ compound. Pd ion can be used a suitable tuning parameter to explore the intricate balance between CDW and superconductivity in 2$H$-TaSe$_2$. A dome shape $T_{\rm{C}}(x)$ curve is observed with an optimal $T_{\rm{C}}\sim$ 3.3 K for $x$ = 0.08-0.09, constituting enhancement of $T_{\rm{C}}$ of 2$H$-TaSe$_2$ by a factor of 24. Pd intercalation drastically suppresses the CCDW phase and but does not have any strong effect on the ICDW transition in 2$H$-TaSe$_2$. Analysis of specific heat data shows an evidence that Pd intercalation increase the effective electron-phonon coupling and density of states at the Fermi energy. Observation of qualitative proportionality between $N(E_{\rm{F}})$ and $T_{\rm{C}}$ in Pd$_x$TaSe$_2$ with the same structure suggests that the increase of $N(E_{\rm{F}})$ (or $\lambda_{ep}$) becomes more necessary to increase $T_{\rm{C}}$. The temperature-dependence of $H_{c2}$, $C_{el}$ and $\gamma_H$ all confirm the presence of multi-gap superconductivity in the 2$H$-Pd$_x$TaSe$_2$ compound.
\begin{table}[h]
  \centering
  \caption{Comparison of physical parameters ($T_{\rm{C}}$, $\gamma_n$, and $\theta_D$) and calculated density of states ($N(E_{\rm{F}})$) and electron-phonon constants ($\lambda_{ep}$) in various TMDCs. $^\ast$ denote the values of parameters which we have calculated using equation (\ref{ep}) from the given parameters given in ref.\cite{wagner}.}\label{scp}
\begin{tabular}{|C{3cm}|C{1cm}|C{2.5cm}|C{1.2cm}|C{1cm}|C{2.5cm}|L{2cm}|}
  \hline
  &$T_{\rm{C}}$&$\gamma_n$&$\theta_D$&&$N(E_{\rm{F}})$ &\\[-1ex]
  \raisebox{1.2ex}{Compound}&(K)&(mJ/mol K$^{-2}$)&(K)&\raisebox{1.2ex}{$\lambda_{ep}$}&states/eV/f.u.&\raisebox{1.2ex}{Ref.} \\
  \hline                     2$H$-TaSe$_2$&0.14&5&202&0.4&1.51&this work\\
  \textbf{2$H$-Pd$_{0.09}$TaSe$_2$}&\textbf{3.3}&\textbf{8.5}&\textbf{196}&\textbf{0.67}&\textbf{2.16}&this work\\
  2$H$-Ni$_{0.02}$TaSe$_2$&2.7&8.13&196&0.65&2.09&\cite{li}\\
  3$R$-TaSe$_{2-x}$Te$_x$&2.4&7.25&184&0.64&1.88&\cite{luo1}\\
  3$R$-Ta$_{0.9}$Mo$_{0.1}$Se$_2$&2.2&7.27&203&0.61&1.92&\cite{luo2}\\
  3$R$-Ta$_{0.9}$W$_{0.1}$Se$_2$&2&5.83&198&0.59&1.56&\cite{luo2}\\
  1$T$-TaSe$_{2-x}$Te$_x$&0.4&2.91&147&0.51&0.82&\cite{luo1}\\
  \hline
  2$H$-TaS$_2$ &0.8&8&165&0.51$^\ast$&2.2$^\ast$& \cite{wagner}\\
  2$H$-Cu$_{0.03}$TaS$_2$&4.5&12&$\sim$165&0.72$^\ast$&2.8$^\ast$& \cite{wagner}\\
  \hline
  1$T$-Cu$_{0.08}$TiSe$_2$& 4.1& 4.3& - & 0.8 & 1.8&\cite{mora1}\\
  \hline
\end{tabular}
\end{table}
\section*{Methods}
Polycrystalline samples of Pd$_x$TaSe$_2$ for different values of $x$ were synthesized via solid state reaction method. Stoichiometric amounts of Pd (99.98\%), Ta (99.9\%) and Se (99.999\%) powder were mixed, grounded well and pelletized. The pellets were sealed in an evacuated quartz tube and then kept at 900 $^{\circ}$C for 4 days. The obtained pellets were crushed, regrounded, pelletized again and resealed in an evacuated quartz tube. The pellets were reannealed for another 5 days at 900 $^{\circ}$C. Single crystal of the Pd$_x$TaSe$_2$ were grown by the chemical vapor transport method using SeCl$_4$ as a transport agent. Millimeter-sized hexagonal shaped crystals were collected from the cold end of the quartz tube. The phase purity of the samples were determined by the powder x-ray diffraction using Cu K$_{\alpha}$ radiation at room temperature. Rietveld analyses of the powder diffraction pattern were performed using the fullprof suits. Resistivity was measured by the conventional four probe method using silver paste (Dupont 4929N$^{\rm{TM}}$) in a PPMS$^{\rm TM}$ (Quantum Design) and a dilution refrigerator. Dc magnetic susceptibility was measured in a MPMS$^{\rm TM}$ (Quantum design). Specific heat data were measured with PPMS$^{\rm{TM}}$ and a dilution refrigerator by using a custom-made program employing the relaxation method with two relaxation times. To determine the actual composition of Pd in the samples, electron probe micro analysis (EPMA) was performed on the polycrystalline samples by considering the Pd metal as standard specimens.

\section*{Acknowledgements}
This work was financially supported by the National Creative Research Initiative (2010-0018300) through the NRF of Korea. B. H. Min also partially supported by NRF-2015R1D1A4A01019960 and Brain Korea 21 plus project.
\end{document}